\documentclass[11pt]{article}

\usepackage[margin=1in]{geometry}
\usepackage{microtype}

\usepackage[numbers,sort&compress]{natbib}

\usepackage[colorlinks=true, citecolor=blue, linkcolor=blue, urlcolor=blue]{hyperref}

\usepackage{amsmath,amssymb,amsthm}
\usepackage{booktabs}
\usepackage{algorithm}
\usepackage{algorithmic}

\usepackage{graphicx}
\usepackage{caption}
\usepackage{authblk}
\title{DiffKnock: Diffusion-based Knockoff Statistics for Neural Networks Inference}

\author{Heng Ge \quad Qing Lu}
\affil{Department of Biostatistics, University of Florida \\
\texttt{heng.ge@ufl.edu, lucienq@ufl.edu}}

\date{}

\begin{document}
\maketitle

\begin{abstract}
We introduce \emph{DiffKnock}, a diffusion-based knockoff framework for high-dimensional feature selection with finite-sample false discovery rate (FDR) control. DiffKnock addresses two key limitations of existing knockoff methods: preserving complex feature dependencies and detecting non-linear associations. Our approach trains diffusion models to generate valid knockoffs and uses neural network–based gradient and filter statistics to construct antisymmetric feature importance measures. Through simulations, we showed that DiffKnock achieved higher power than autoencoder-based knockoffs while maintaining target FDR, indicating its superior performance in scenarios involving complex non-linear architectures. Applied to murine single-cell RNA-seq data of LPS-stimulated macrophages, DiffKnock identifies canonical NF-$\kappa$B target genes (\textit{Ccl3}, \textit{Hmox1}) and regulators (\textit{Fosb}, \textit{Pdgfb}). These results highlight that, by combining the flexibility of deep generative models with rigorous statistical guarantees, DiffKnock is a powerful and reliable tool for analyzing single-cell RNA-seq data, as well as high-dimensional and structured data in other domains.
\end{abstract}

\section{Introduction}

High-dimensional problems present fundamental challenges for statistical inference, especially when dealing with complex feature dependencies and nonlinear relationships \citep{buhlmann2011statistics,o2016statistical}. This challenge spans genomics \citep{he2011variable,stuart2019integrative}, neuroscience \citep{vu2011encoding,jenatton2012multiscale}, and other domains where identifying truly relevant features while controlling false discoveries becomes increasingly difficult with complex feature dependencies \citep{pudjihartono2022review}.

Traditional multiple testing approaches, such as the Benjamini--Hochberg procedure \citep{benjamini1995controlling}, often fail in high-dimensional settings with complex correlations and non-linear relationships \citep{leek2008general,fan2012estimating}. The violation of their distributional assumptions frequently leads to inflated false discovery rates or reduced power \citep{efron2007correlation,sun2009large}.

The model-X knockoff framework \citep{barber2015controlling,candes2018panning} offers a fundamentally different approach. By constructing synthetic knockoff variables that mimic feature correlations while remaining conditionally independent of the response, it achieves finite-sample FDR control without assumptions about the conditional distribution. However, practical application faces two challenges: generating valid knockoffs preserving complex dependencies, and computing powerful statistics for non-linear relationships while maintaining the required antisymmetric structure.

Deep learning provides a promising solution, with generative models successfully learning high-dimensional distributions \citep{goodfellow2014generative,creswell2018generative} and neural networks capturing non-linear relationships \citep{lecun2015deep,GoodBengCour16}. DeepLINK \citep{lu2018deeppink,zhu2021deeplink} pioneered the integration of deep learning with knockoff inference, using variational autoencoders and filter-based statistics to improve power over linear approaches. However, VAEs' Gaussian latent space assumptions and reconstruction objectives can lead to over-smoothing that fails to preserve fine-grained correlations.

To address the limitations of existing knockoff methods, we present DiffKnock, leveraging diffusion models for knockoff generation while maintaining FDR control. Diffusion models \citep{ho2020denoising,sohl2015deep} decompose generation into simple denoising steps, capturing complex dependencies with greater stability than GANs and having fewer distributional assumptions than VAEs. Our contributions are: (1) the development of a new diffusion-based knockoff generation method using transformer architecture to preserve complex correlation structures; (2) the adoption of innovative gradient-based and filter-based neural network statistics maintaining antisymmetric properties for FDR control; (3) demonstration through simulations and single-cell RNA-seq analysis that DiffKnock achieves superior power while maintaining FDR control, particularly for complex non-linear relationships. An anonymized implementation of our method is provided in the supplementary material.

\section{Methodologies}

This section develops the technical foundations of DiffKnock, integrating diffusion-based generative modeling \citep{sohl2015deep,ho2020denoising} with the model-X knockoff framework \citep{barber2015controlling,candes2018panning} for controlled feature selection. We describe our approach in four components: the statistical guarantees provided by the knockoff framework (Section 2.1), neural network architectures for computing antisymmetric test statistics (Section 2.2), our diffusion model design for generating valid knockoffs (Section 2.3), and the complete DiffKnock algorithm implementation with computational complexity (Section 2.4). 

\subsection{The Model-X Knockoff Framework}

The model-X knockoff framework provides principled false discovery rate (FDR) control for feature selection without assumptions about the conditional distribution $P(\mathbf{y}|\mathbf{X})$. For features $\mathbf{X} = [\mathbf{x}_1, \ldots, \mathbf{x}_p] \in \mathbb{R}^{n \times p}$ and response $\mathbf{y} \in \mathbb{R}^n$, the framework constructs knockoff variables $\tilde{\mathbf{X}} = [\tilde{\mathbf{x}}_1, \ldots, \tilde{\mathbf{x}}_p] \in \mathbb{R}^{n \times p}$ satisfying two properties:

\textbf{Property 1 (Exchangeability):} For any subset $S \subseteq \{1, \ldots, p\}$, swapping original and knockoff features in $S$ preserves the joint distribution:
\begin{equation}
(\mathbf{X}, \tilde{\mathbf{X}})_{\text{swap}(S)} \stackrel{d}{=} (\mathbf{X}, \tilde{\mathbf{X}})
\end{equation}

\textbf{Property 2 (Conditional Independence):} Knockoffs contain no additional information about the response:
\begin{equation}
\tilde{\mathbf{X}} \perp \mathbf{y} | \mathbf{X}
\end{equation}

Given valid knockoffs, we compute antisymmetric statistics $W_j$ for each feature $j \in \{1, \ldots, p\}$ that measure the difference in importance between $\mathbf{x}_j$ and $\tilde{\mathbf{x}}_j$. The knockoff+ procedure then selects features by finding an adaptive threshold:
\begin{equation}
\tau = \min\left\{t > 0 : \frac{1 + |\{j: W_j \leq -t\}|}{\max(|\{j: W_j \geq t\}|, 1)} \leq q\right\}
\end{equation}
where $q \in (0,1)$ is the target FDR level. The selected features $\hat{S} = \{j : W_j \geq \tau\}$ satisfy:
\begin{equation}
\text{FDR} = \mathbb{E}\left[\frac{|\hat{S} \cap S_0^c|}{\max(|\hat{S}|, 1)}\right] \leq q
\end{equation}
where $S_0 \subseteq \{1, \ldots, p\}$ denotes truly associated features and $S_0^c$ the set of null (non-associated) features. This guarantee holds in finite samples without distributional assumptions, making it particularly attractive for applications where sample sizes are limited and the underlying relationships are complex.

\subsection{Neural Network Architecture for Knockoff Statistics}

Traditional knockoff methods have primarily been used in conjunction with linear models or simple test statistics. However,  high-dimensional datasets often exhibit intricate non-linear dependencies that require more advanced tools \citep{lecun2015deep}. The Model-X knockoff framework is particularly well-suited for this setting, as it can be seamlessly integrated with neural network architectures \citep{lu2018deeppink,zhu2021deeplink}. By jointly processing original and knockoff features within a carefully designed network that enforces the required antisymmetric structure, we can draw valid inferences while capturing the complex patterns present in the data.

The network begins with a pairwise filter layer that processes each feature-knockoff pair $(\mathbf{x}_j, \tilde{\mathbf{x}}_j)$ through learnable weights $(z_j, \tilde{z}_j) \in \mathbb{R}^2$:
\begin{equation}
\mathbf{f}_j = \frac{z_j}{|z_j| + |\tilde{z}_j|} \cdot \mathbf{x}_j + \frac{\tilde{z}_j}{|z_j| + |\tilde{z}_j|} \cdot \tilde{\mathbf{x}}_j
\end{equation}
This normalization ensures unit sum while allowing the network to learn relative importance. The filtered features pass through a multi-layer perceptron:
\begin{equation}
\mathbf{h}^{(l)} = \sigma(\mathbf{W}^{(l)}\text{LayerNorm}(\mathbf{h}^{(l-1)}) + \mathbf{b}^{(l)})
\end{equation}
where $\mathbf{W}^{(l)}$ and $\mathbf{b}^{(l)}$ are the weight matrix and bias vector for layer $l$, and $\sigma(\cdot)$ denotes the activation function. Layer normalization provides stability for features with varying scales, crucial when features span multiple orders of magnitude.

From the trained network, we extract two types of knockoff statistics, for which we prove the required antisymmetry property in the supplementary material:

\textbf{Gradient-based Statistics:} Measure feature importance through loss sensitivity:
\begin{equation}
W_j^{\text{grad}} = \frac{1}{n}\sum_{i=1}^n \left|\frac{\partial \mathcal{L}(y_i, \hat{y}_i)}{\partial x_{ij}}\right| - \left|\frac{\partial \mathcal{L}(y_i, \hat{y}_i)}{\partial \tilde{x}_{ij}}\right|
\end{equation}
where $x_{ij}$ and $\tilde{x}_{ij}$ are the $i$-th sample values of the $j$-th original and knockoff features, $y_i$ is the true response, $\hat{y}_i$ is the predicted response, and $\mathcal{L}$ is the loss function. The gradient of the loss with respect to an input feature measures how sensitively the model’s prediction error responds to infinitesimal perturbations of that feature \citep{simonyan2013deep,sundararajan2017axiomatic,shrikumar2017learning}. Features with large gradient magnitudes contribute more strongly to predictive performance, while near-zero gradients indicate limited influence. This property makes loss gradients a natural measure of feature importance, and by contrasting gradients between original and knockoff variables we obtain valid antisymmetric statistics for knockoff inference.

\textbf{Filter-based Statistics:} 
We define the statistic for feature $j$ as
\begin{equation}
W_j^{\text{filter}} = (w_j^{\text{eff}} \cdot z_j)^2 - (w_j^{\text{eff}} \cdot \tilde{z}_j)^2
\end{equation}
where $z_j$ and $\tilde{z}_j$ are the filter weights for the original and knockoff features, and $w_j^{\text{eff}} \in \mathbb{R}$ is the effective input weight obtained by multiplying through all linear transformations from the input to the output layer. Intuitively, $w_j^{\text{eff}}$ summarizes how strongly information from feature $j$ propagates through the network, while the filter weights $z_j$ and $\tilde{z}_j$ determine the relative allocation between the original and its knockoff. 
If $(w_j^{\text{eff}} z_j)^2$ greatly exceeds $(w_j^{\text{eff}} \tilde{z}_j)^2$, the model relies more heavily on the original feature, yielding a large positive statistic. This construction provides a direct, parameter-based measure of importance \citep{lu2018deeppink,zhu2021deeplink}.

\subsection{Diffusion Models for Knockoff Generation}

We propose to use diffusion models for knockoff generation, which is motivated by their stability in high-dimensional settings and ability to capture complex distributions. Diffusion models decompose generation into gradual denoising steps, avoiding the optimization challenges of GANs while providing more flexibility than VAEs.

The forward diffusion process adds noise over $T$ timesteps:
\begin{equation}
q(\mathbf{x}_t | \mathbf{x}_{t-1}) = \mathcal{N}(\mathbf{x}_t; \sqrt{1-\beta_t}\mathbf{x}_{t-1}, \beta_t\mathbf{I})
\end{equation}
with variance schedule $\{\beta_t\}_{t=1}^T$ where $\beta_t \in (0,1)$. Through reparameterization:
\begin{equation}
q(\mathbf{x}_t | \mathbf{x}_0) = \mathcal{N}(\mathbf{x}_t; \sqrt{\bar{\alpha}_t}\mathbf{x}_0, (1-\bar{\alpha}_t)\mathbf{I})
\end{equation}
where $\alpha_t = 1 - \beta_t$ and $\bar{\alpha}_t = \prod_{s=1}^t \alpha_s$.

The reverse process learns to denoise using a neural network $\boldsymbol{\epsilon}_\theta(\mathbf{x}_t, t)$ that predicts noise given corrupted data and timestep. We employ a transformer architecture (see Section 2.4) that captures long-range feature interactions through self-attention mechanisms. The model is trained to minimize:
\begin{equation}
\mathcal{L} = \mathbb{E}_{t, \mathbf{x}_0, \boldsymbol{\epsilon}} \left[ \|\boldsymbol{\epsilon} - \boldsymbol{\epsilon}_\theta(\sqrt{\bar{\alpha}_t}\mathbf{x}_0 + \sqrt{1-\bar{\alpha}_t}\boldsymbol{\epsilon}, t)\|^2 \right]
\end{equation}
where $\boldsymbol{\epsilon} \sim \mathcal{N}(\mathbf{0}, \mathbf{I})$. We adopt a cosine noise schedule \citep{nichol2021improved}:
\begin{equation}
\bar{\alpha}_t = \frac{f(t)}{f(0)}, \quad f(t) = \cos\left(\frac{t/T + s}{1 + s} \cdot \frac{\pi}{2}\right)^2
\end{equation}
where $s > 0$ is a small offset parameter. This maintains consistent signal-to-noise ratios, crucial for preserving feature patterns across scales.

To generate knockoffs, we start from noise $\mathbf{x}_T \sim \mathcal{N}(\mathbf{0}, \mathbf{I})$ and iteratively denoise:
\begin{equation}
p_\theta(\mathbf{x}_{t-1} | \mathbf{x}_t) = \mathcal{N}(\mathbf{x}_{t-1}; \boldsymbol{\mu}_\theta(\mathbf{x}_t, t), \sigma_t^2\mathbf{I})
\end{equation}
where the mean is computed as:
\begin{equation}
\boldsymbol{\mu}_\theta(\mathbf{x}_t, t) = \frac{1}{\sqrt{\alpha_t}}\left(\mathbf{x}_t - \frac{\beta_t}{\sqrt{1-\bar{\alpha}_t}}\boldsymbol{\epsilon}_\theta(\mathbf{x}_t, t)\right)
\end{equation}
and $\sigma_t^2$ is the variance of the reverse process.

The resulting knockoffs are generated by iterative denoising, starting from Gaussian noise and guided by the transformer-based diffusion model. This procedure captures both local and global feature dependencies while maintaining stability in high-dimensional settings. To ensure approximate validity under the Model-X framework, we further apply marginal-matching corrections (see Section 2.4), which enforce distributional consistency between original and knockoff features. Together, these steps yield knockoffs that preserve complex dependency structures while respecting exchangeability and remaining conditionally independent of the response. Formal proofs of conditional independence, exchangeability, and approximate FDR control are provided in the Supplementary Material.

\subsection{The DiffKnock Algorithm Implementation and Computational Complexity}

\begin{algorithm}[htbp]
\caption{DiffKnock: Diffusion-based Knockoff Generation for Feature Selection}
\begin{algorithmic}[1]
\STATE \textbf{Input:} Feature matrix $\mathbf{X} \in \mathbb{R}^{n \times p}$, response $\mathbf{y} \in \mathbb{R}^n$, target FDR level $q \in (0,1)$
\STATE \textbf{Output:} Selected feature set $\hat{S} \subseteq \{1, \ldots, p\}$
\STATE
\STATE \textbf{Stage 1: Data Preprocessing}
\STATE \hspace{0.5cm} Apply domain-appropriate transformation (e.g., log for positive-valued data)
\STATE \hspace{0.5cm} Normalize features to stable range for diffusion training
\STATE
\STATE \textbf{Stage 2: Diffusion Model Training}
\STATE \hspace{0.5cm} Initialize transformer-based diffusion model with parameters $\theta$
\STATE \hspace{0.5cm} \textbf{for} epoch $= 1$ to $N_{\text{epochs}}$ \textbf{do}
\STATE \hspace{1cm} Sample batch $\mathbf{X}_{\text{batch}}$ from normalized data
\STATE \hspace{1cm} Sample timesteps $t \sim \text{Uniform}(1, T)$
\STATE \hspace{1cm} Sample noise $\boldsymbol{\epsilon} \sim \mathcal{N}(\mathbf{0}, \mathbf{I})$
\STATE \hspace{1cm} Compute loss: $\mathcal{L} = \|\boldsymbol{\epsilon} - \boldsymbol{\epsilon}_\theta(\sqrt{\bar{\alpha}_t}\mathbf{X}_{\text{batch}} + \sqrt{1-\bar{\alpha}_t}\boldsymbol{\epsilon}, t)\|^2$
\STATE \hspace{1cm} Update parameters $\theta$ via gradient descent
\STATE \hspace{0.5cm} \textbf{end for}
\STATE
\STATE \textbf{Stage 3: Knockoff Generation}
\STATE \hspace{0.5cm} Sample initial noise: $\mathbf{X}_T \sim \mathcal{N}(\mathbf{0}, \mathbf{I})$
\STATE \hspace{0.5cm} \textbf{for} $t = T$ down to $1$ \textbf{do}
\STATE \hspace{1cm} Predict noise: $\hat{\boldsymbol{\epsilon}} = \boldsymbol{\epsilon}_\theta(\mathbf{X}_t, t)$
\STATE \hspace{1cm} Compute mean: $\boldsymbol{\mu}_t = \frac{1}{\sqrt{\alpha_t}}(\mathbf{X}_t - \frac{\beta_t}{\sqrt{1-\bar{\alpha}_t}}\hat{\boldsymbol{\epsilon}})$
\STATE \hspace{1cm} Sample: $\mathbf{X}_{t-1} \sim \mathcal{N}(\boldsymbol{\mu}_t, \sigma_t^2\mathbf{I})$
\STATE \hspace{0.5cm} \textbf{end for}
\STATE \hspace{0.5cm} Apply marginal matching to ensure distributional consistency: $\tilde{\mathbf{X}} = \text{MatchMarginals}(\mathbf{X}, \mathbf{X}_0)$
\STATE
\STATE \textbf{Stage 4: Feature Selection with FDR Control}
\STATE \hspace{0.5cm} Train neural network on augmented features $[\mathbf{X}, \tilde{\mathbf{X}}]$
\STATE \hspace{0.5cm} Compute knockoff statistics $W_j$ for $j = 1, \ldots, p$
\STATE \hspace{0.5cm} Find threshold: $\tau = \min\{t > 0 : \frac{1 + |\{j: W_j \leq -t\}|}{\max(|\{j: W_j \geq t\}|, 1)} \leq q\}$
\STATE \hspace{0.5cm} Select features: $\hat{S} = \{j : W_j \geq \tau\}$
\STATE \hspace{0.5cm} \textbf{return} $\hat{S}$
\end{algorithmic}
\end{algorithm}

The implementation of DiffKnock requires careful consideration of computational efficiency and numerical stability across its four stages. Following data preprocessing with domain-appropriate transformations, the diffusion model training phase represents the primary computational bottleneck with complexity $\mathcal{O}(N_{\text{epochs}} \cdot n \cdot T \cdot (p^2 \cdot L \cdot h + p \cdot d^2))$, where $n$ is the number of samples, $p$ is the number of features, $T$ is the number of diffusion timesteps, $L$ denotes transformer layers, $h$ represents attention heads, and $d$ is the hidden dimension. Our transformer architecture employs conditional layer normalization that modulates the scale and shift parameters based on the timestep embedding: $\text{CLN}(\mathbf{x}, t) = \gamma(t) \odot \frac{\mathbf{x} - \mu}{\sigma} + \beta(t)$, where $\mu$ and $\sigma$ are the mean and standard deviation of $\mathbf{x}$, $\odot$ denotes element-wise multiplication, and $\gamma(t), \beta(t) \in \mathbb{R}^d$ are learned projections from a time embedding network consisting of sinusoidal encodings followed by two linear layers with GELU activation.

The knockoff generation phase implements DDPM sampling with computational complexity $\mathcal{O}(T \cdot p^2 \cdot L \cdot h)$ per sample. Critical to maintaining valid knockoffs, we employ a marginal matching procedure that preserves empirical distributions while retaining learned dependencies. For each feature $j \in \{1, \ldots, p\}$, we compute the empirical quantile function $\hat{F}_j^{-1}$ of the original data and apply it to the uniform ranks of the generated samples: $\tilde{x}_{ij} = \hat{F}_j^{-1}(\text{rank}(\tilde{x}_{ij}^{(0)})/n)$, where $\tilde{x}_{ij}^{(0)}$ denotes the $i$-th sample of the $j$-th feature from the raw diffusion output. This operation requires $\mathcal{O}(np \log n)$ time to sort the operations across all features.

For computing knockoff statistics, we implement both gradient-based and filter-based approaches described in Section 2.2. The gradient-based method computes importance scores via backpropagation: $W_j^{\text{grad}} = \frac{1}{n}\sum_{i=1}^n |\nabla_{x_{ij}} \mathcal{L}(y_i, \hat{y}_i)| - \frac{1}{n}\sum_{i=1}^n |\nabla_{\tilde{x}_{ij}} \mathcal{L}(y_i, \hat{y}_i)|$, where $\mathcal{L}$ is the loss function, $\hat{y}_i = f_\theta([\mathbf{x}_i, \tilde{\mathbf{x}}_i])$ is the network prediction, and $f_\theta$ denotes the trained neural network with parameters $\theta$. The filter-based approach computes statistics as $W_j^{\text{filter}} = (z_j \cdot w_j^{\text{eff}})^2 - (\tilde{z}_j \cdot w_j^{\text{eff}})^2$, where $(z_j, \tilde{z}_j)$ are the normalized filter weights and $w_j^{\text{eff}}$ represents the effective weight obtained by multiplying through the network layers as described. Both approaches require $\mathcal{O}(n \cdot p \cdot H)$ operations, where $H$ is the total number of hidden units across layers.

The FDR control procedure implements the knockoff+ filter, iterating through at most $p$ unique thresholds to find the smallest $\tau$ satisfying the FDR constraint. This requires $\mathcal{O}(p \log p)$ time to sort the statistics. Memory requirements are dominated by the transformer model with $\mathcal{O}(L \cdot (p^2 \cdot h + p \cdot d^2))$ parameters and the storage of both original and knockoff features requiring $\mathcal{O}(2np)$ space. The complete algorithm scales as $\mathcal{O}(N_{\text{epochs}} \cdot n \cdot T \cdot p^2 \cdot L \cdot h)$ in time and $\mathcal{O}(np + L \cdot p^2 \cdot h)$ in space, making it tractable for datasets with thousands of features when using GPU acceleration.

\section{Simulation Studies}

\subsection{Simulation Design}

We evaluated DiffKnock through comprehensive simulations that mimic the characteristics of RNA-seq transcripts per million (TPM) data. In the simulation, we generated $n = 1000$ samples with $p = 50$ genes, where $s = 5$ genes are causally related to the outcome. This setup reflects typical genomic studies where the number of features is moderate and only a small subset influences the phenotype.

The data generation process incorporates realistic features of gene expression data. Library sizes are drawn from a log-normal distribution $L_i \sim \text{LogNormal}(\log(10^6), 0.5)$ for $i = 1, \ldots, n$ to model sequencing depth variation. Gene expression exhibits block correlation structure to mimic co-expression modules:
\begin{equation}
\boldsymbol{\Sigma}_{\text{block}} = (1 - \rho)\mathbf{I} + \rho\mathbf{1}\mathbf{1}^T, \quad \rho \sim \text{Uniform}(0.4, 0.8)
\end{equation}
where $\mathbf{I} \in \mathbb{R}^{p \times p}$ is the identity matrix and $\mathbf{1} \in \mathbb{R}^p$ is the vector of ones. Baseline expression levels span several orders of magnitude with $\log(\text{expression}_j) \sim \text{Uniform}(2, 6)$ for $j = 1, \ldots, p$. Causal genes are preferentially selected from highly expressed genes: $P(\text{gene } j \text{ is causal}) \propto \exp(\bar{x}_j / \max_k \bar{x}_k)$ where $\bar{x}_j$ is the mean expression of gene $j$.

To comprehensively evaluate methods' performance, we test eight distinct feature-outcome relationships with varying complexity. For each scenario, we evaluate performance across signal amplitudes $A \in [0.5, 7.0]$ using 20 evenly spaced values, with 50 independent simulations per configuration at the target FDR level $q = 0.2$. The base outcome is computed as $\mathbf{y}_{\text{base}} = \mathbf{X}\boldsymbol{\beta}$, where $\beta_j \sim \mathcal{N}(0, A)$ for $j \in S_{\text{causal}}$ and $\beta_j = 0$ otherwise. Each scenario applies different transformations:

\textbf{Linear:} Baseline with additive effects: $\mathbf{y} = \mathbf{y}_{\text{base}} + \boldsymbol{\epsilon}$.

\textbf{Polynomial:} Higher-order terms favoring gradient methods: $\mathbf{y} = \mathbf{y}_{\text{base}} + 0.3\mathbf{y}_{\text{base}}^2 + 0.1\mathbf{y}_{\text{base}}^3 + \sum_{j \in S_{\text{causal}}^{(2)}} 0.2\mathbf{x}_j^2 + \boldsymbol{\epsilon}$ where $S_{\text{causal}}^{(2)}$ denotes the first two causal genes.

\textbf{Mixed:} Combination of linear and nonlinear components: $\mathbf{y} = 0.3\mathbf{y}_{\text{base}} + 0.3\tanh(\mathbf{y}_{\text{base}}) + 0.2(\mathbf{y}_{\text{base}}^2 - \mathbb{E}[\mathbf{y}_{\text{base}}^2]) + 0.2(\exp(\text{clip}(0.3\mathbf{y}_{\text{base}}, -5, 5)) - 1) + \boldsymbol{\epsilon}$.

\textbf{Information Bottleneck:} Compression followed by gene-specific expansion:
$\mathbf{y} = \tanh(\mathbf{y}_{\text{base}}/2) + \sum_{j \in S_{\text{causal}}} 0.3 f_j(\tanh(\mathbf{y}_{\text{base}}/2)) \odot \mathbf{x}_j + 0.2(\tanh(\mathbf{y}_{\text{base}}/2) - \mathbf{y}_{\text{base}})^2 + \boldsymbol{\epsilon}$ where $f_j \in \{\exp(-|\cdot|), (\cdot)^2\text{sign}(\cdot), \sin(\pi\cdot)\}$ are gene-specific expansion functions and $\odot$ denotes element-wise multiplication.

\textbf{Multiscale Frequency:} Multiple oscillatory components are introduced at different frequencies ($0.5, 1, 2, 4, 8$) for the causal genes, combined with higher-frequency details and pairwise frequency mixing. The overall signal is modulated by an envelope $1+0.5\tanh(\mathbf{y}_{\text{base}}/2)$ and added to $\mathbf{y}_{\text{base}}$ with noise.

\textbf{Network Propagation:} Direct signals from causal genes are propagated through three transformation layers: (i) a $\tanh$ nonlinearity, (ii) a damped exponential modulation, and (iii) a signed square-root scaling. Each layer contributes gene-specific weighted effects, which are aggregated together with the interaction term and noise.

For all scenarios, we add Gaussian noise $\boldsymbol{\epsilon} \sim \mathcal{N}(\mathbf{0}, \mathbf{I})$ and standardize the outcome to zero mean and variance of one. This diverse set of transformations tests the methods' ability to capture relationships ranging from simple linear associations to complex non-linear dependencies with multiple interaction levels.

All simulations and real data experiments were conducted on the University of Florida HiPerGator high-performance computing cluster, using NVIDIA B200 GPUs with CUDA acceleration. The diffusion model training and knockoff generation stages were parallelized across GPUs, while downstream neural network training for knockoff statistics was performed on the same infrastructure. This environment provided sufficient memory and compute throughput to handle up to several hundred features and thousands of samples in our experiments.

\subsection{Implementation and Baseline Methods}

We compare three methods for different signal amplitudes ranging from 0.5 to 7.0, evaluated over 50 independent simulations per configuration with the target FDR level of $q = 0.2$:

\textbf{DiffKnock with Gradient-based Statistics:} Our proposed method uses diffusion models for knockoff generation and gradient-based importance measures. The diffusion model employs a 6-layer transformer with 256-dimensional hidden states, 8 attention heads, and 1000 timesteps with cosine scheduling. Gradients are computed via backpropagation through the trained neural network to capture the local importance of features.

\textbf{DiffKnock with Filter-based Statistics:} Uses the same diffusion-generated knockoffs but extracts importance from the learned filter weights: $W_j = (w_j^{\text{eff}} \cdot z_j)^2 - (w_j^{\text{eff}} \cdot \tilde{z}_j)^2$, where $w_j^{\text{eff}}$ represents effective weights propagated through network layers.

\textbf{DeepLINK (Baseline):} The original DeepLINK \citep{lu2018deeppink,zhu2021deeplink} method using autoencoder-based knockoff generation with filter-based statistics. The autoencoder uses a bottleneck dimension of 3 with ELU activations, trained for 300 epochs. The feature selection network aligns with our architecture for a fair comparison.

All methods use identical neural network architectures for feature selection: two hidden layers [50, 20] with layer normalization, ReLU activation, and dropout rate 0.1, trained for 1000 epochs with Adam optimizer.

\subsection{Results and Analysis}
\begin{figure*}[t]
  \centering
  \vspace{.3in}
  \includegraphics[width=\textwidth]{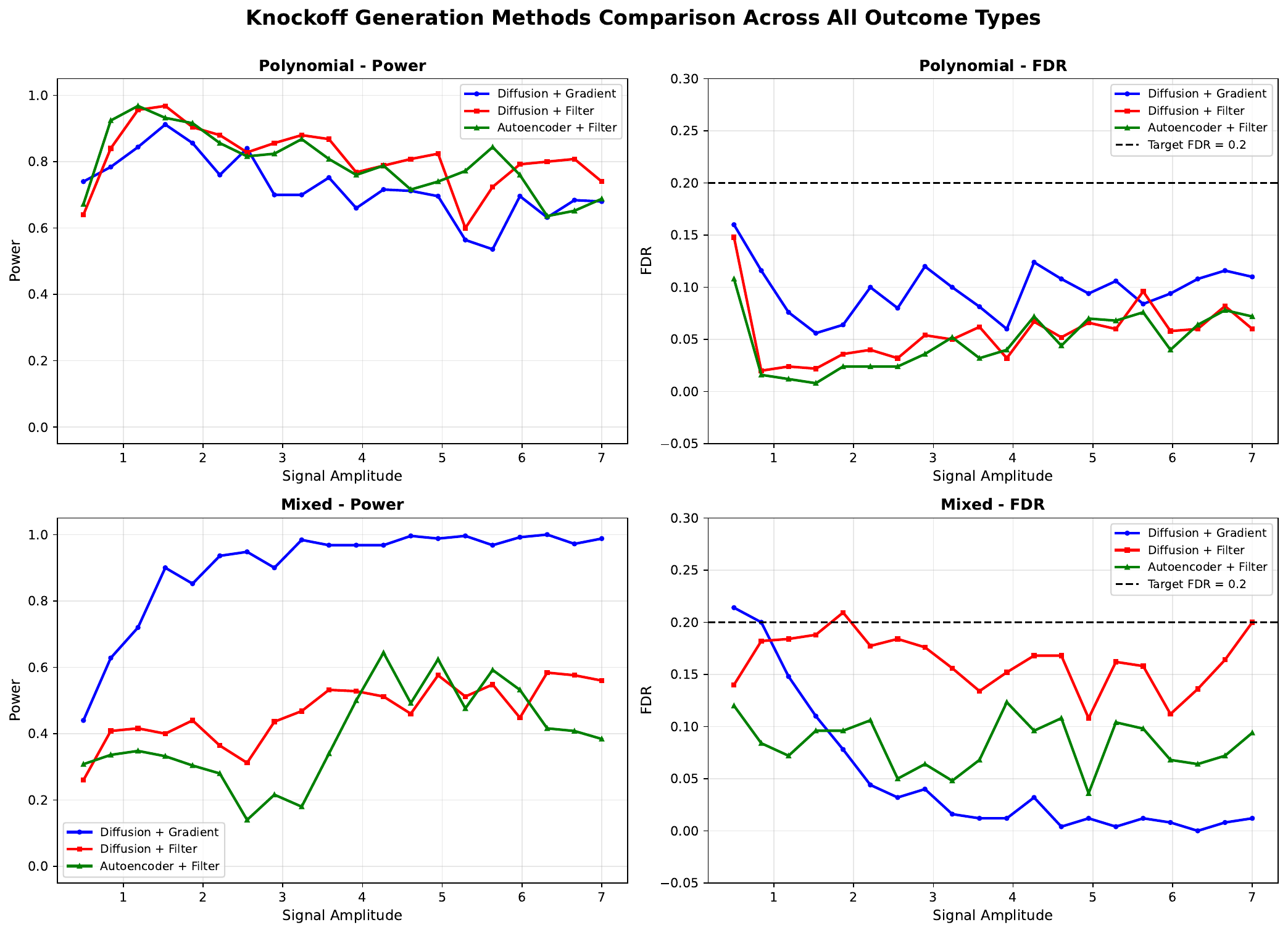}
  \vspace{.3in}
  \caption{Power and FDR for Polynomial and Mixed Scenario.}
  \label{Simulation1}
\end{figure*}

Figure \ref{Simulation1} presents power and FDR across signal amplitudes for the polynomial and mixed scenarios. These representative cases demonstrate the robustness of our approach across diverse non-linear relationships:

\textbf{Polynomial Scenario:} Filter-based statistics consistently outperform gradient-based approaches across all amplitudes, with both diffusion and autoencoder methods achieving power near 0.8 at higher signal strengths. The diffusion-based approach shows marginally better performance, particularly in the mid-range amplitudes (2.0-4.0) and high-range amplitudes (6.0-7.0), while maintaining comparable FDR control. The superiority of filter-based statistics in this setting aligns with the polynomial structure where squared terms ($0.2\mathbf{x}_j^2$ for $j \in S_{\text{causal}}^{(2)}$) create stable, global nonlinear effects that are well captured by the effective weights $(w_j^{\text{eff}} \cdot z_j)^2$. The gradient-based approach, which measures local sensitivities, appears less suited to these consistent polynomial transformations.

\textbf{Mixed Scenario:} A striking performance separation emerges, with gradient-based statistics (using diffusion knockoffs) achieving near-perfect power (approaching 1.0) for amplitudes above 3.0, while filter-based methods attain around 0.4-0.6 power. The mixed outcome combines multiple non-linear transformations, including $\tanh$, squared, and exponential components, creating a complex response surface where feature importance varies locally. Gradient-based statistics excel here by capturing these position-dependent sensitivities through $\frac{\partial \mathcal{L}}{\partial x_{ij}}$, adapting to how each transformation component contributes differently across the input space.

Diffusion-based knockoff generation demonstrates great FDR control in both scenarios. While the autoencoder approach shows comparable or occasionally better FDR in the polynomial case, the diffusion model maintains more consistent control near the target threshold of 0.2, particularly evident in the mixed scenario where gradient-based statistics achieve exceptional power with FDR below 0.05 after amplitude 3.0.

The performance patterns reveal an important insight: the choice of test statistic (gradient vs. filter) should align with the underlying signal structure. Filter-based statistics perform well when non-linearities are global and consistent (polynomial), while gradient-based statistics excel when effects are locally varying (mixed). The diffusion model's ability to generate high-quality knockoffs provides a slight but consistent advantage across both settings, suggesting its broader applicability for capturing the complex correlation structures inherent in gene expression data.

These results underscore that no single approach dominates universally—the optimal method depends on the nature of the underlying biological relationships. The flexibility to employ different statistics with diffusion-generated knockoffs enables adaptation to diverse genomic architectures while maintaining the FDR control.

\section{Real Data Analysis: Murine Single-Cell RNA-Sequencing}

\subsection{Dataset and Preprocessing}

We applied DiffKnock to a murine single-cell RNA-sequencing (scRNA-seq) dataset from Lane et al., investigating the effect of lipopolysaccharides (LPS)-stimulated nuclear factor-$\kappa$B (NF-$\kappa$B) on gene expression \citep{lane2017measuring}. The dataset contains cells under two conditions: unstimulated (202 cells) and LPS-stimulated after 150 minutes (368 cells), providing a binary classification problem relevant to understanding inflammatory response mechanisms.

Following standard scRNA-seq preprocessing guidelines, we filtered cells with mapping rates below 20\% or nonzero expression proportions below 5\%, and removed genes expressed in less than 5\% of cells. The preprocessed data matrix contained TPM expression values for 13,777 genes from 570 cells. Given the computational challenges of applying DiffKnock directly to the whole genome, we employed a screening strategy to reduce dimensionality while preserving signal.

The preprocessing pipeline consisted of three stages:

\textbf{Screening:} We randomly split the 570 cells into three sets: 285 cells (50\%) for screening, 228 cells (40\%) for training, and 57 cells (10\%) for testing. For each of 100 repetitions, we performed distance correlation screening on the screening set to identify genes most associated with the stimulation status. This approach ensures that feature selection is performed independently of the training data, avoiding false postivie results.

\textbf{Normalization:} Selected genes underwent log-transformation with pseudocount ($\log(1 + \text{TPM})$) followed by standardization to zero mean and unit variance, ensuring stable neural network training despite the wide dynamic range of gene expression.

\textbf{Knockoff Generation:} We trained separate diffusion models for each gene set size (20, 30, 50, 100, 200, 500 genes) using the architecture described in Section 3. The transformer-based diffusion model with 6 layers and 256-dimensional hidden states was trained for 300 epochs to learn the complex correlation structure of the selected genes.

\subsection{Results and Biological Interpretation}

We evaluated DiffKnock across different screened gene set sizes, with detailed results for $d = 50$ genes shown in Table~\ref{tab:top_genes}. Both gradient-based and filter-based statistics achieved substantial dimensionality reduction, selecting only 7--8 genes from the initial 50 features while maintaining strong predictive performance, demonstrating effective identification of informative features with FDR control at $q = 0.2$.

\begin{table}[h]
\caption{Top 10 Most Frequently Selected Genes (50 Initial Genes, 100 Repetitions)} \label{tab:top_genes}
\begin{center}
\begin{tabular}{lcc}
\textbf{Gene} & \textbf{Filter (\%)} & \textbf{Gradient (\%)} \\
\hline \\
\textit{Ccl3} & 92.0 & 87.0 \\
\textit{Hmox1} & 78.0 & 71.0 \\
\textit{Cyp51} & 67.0 & 58.0 \\
\textit{Chordc1} & 43.0 & 36.0 \\
\textit{Fosb} & 37.0 & 44.0 \\
\textit{Pdgfb} & 31.0 & 27.0 \\
\textit{Ifrd1} & 18.0 & 19.0 \\
\textit{S100a6} & 12.0 & 9.0 \\
\textit{Tubb6} & 8.0 & 9.0 \\
\textit{Ier3} & 8.0 & 7.0 \\
\end{tabular}
\end{center}
\end{table}

The selected genes demonstrate strong biological relevance to LPS-induced NF-$\kappa$B activation. \textit{Ccl3} (87--92\% selection), encoding macrophage inflammatory protein-1$\alpha$, is a canonical NF-$\kappa$B target that orchestrates chemotaxis of neutrophils and monocytes during acute inflammation \citep{mueller2004ccl3}. \textit{Hmox1} (71--78\% selection) provides cytoprotection through its anti-inflammatory and anti-oxidant activities, and its expression is strongly induced by LPS via NF-$\kappa$B and Nrf2 signaling pathways \citep{paine2010signaling}. \textit{Cyp51}, encoding lanosterol 14$\alpha$-demethylase, reflects metabolic reprogramming during macrophage activation, consistent with evidence that sterol biosynthesis is dynamically regulated during immune responses \citep{blanc2011host}.

Regulatory genes captured by DiffKnock highlight the transcriptional dynamics of the inflammatory cascade. \textit{Fosb}, an immediate-early AP-1 family gene, amplifies inflammatory signaling and contributes to sustained transcriptional responses. \textit{Pdgfb}, encoding platelet-derived growth factor-B, promotes angiogenesis and tissue repair processes in the resolution phase of inflammation \citep{andrae2008role}.

\section{Discussion}
In this paper, we developed a diffusion-based model-X knockoff framework, \textbf{DiffKnock}, that combines deep generative modeling with antisymmetric neural statistics to achieve finite-sample FDR control in high-dimensional and nonlinear regimes. Using a transformer denoiser with post-hoc marginal matching, the method approximates exchangeability while preserving multi-scale dependencies common in genomics. In simulations with linear, polynomial, mixed, information bottleneck, multiscale frequency, and network-propagation settings, DiffKnock delivered strong power with FDR near target. In murine scRNA-seq, it recovered NF-$\kappa$B--related genes (\textit{Ccl3}, \textit{Hmox1}, \textit{Fosb}, \textit{Pdgfb}), showing that diffusion-based knockoffs with neural statistics can detect biologically coherent signals under the FDR control.

An important takeaway from this study is that the choice of statistics should match signal geometry. \textbf{Filter-based statistics}, summarizing parameter pathways, excelled with smooth global nonlinearity (e.g., polynomial terms). \textbf{Gradient-based statistics}, reflecting local loss sensitivity, performed best when effects varied across inputs (e.g., mixed scenario). Since real data rarely fits one regime, we suggest computing both on the same knockoffs, comparing stability, and aggregating via ranks when appropriate. For approximate exchangeability, routine diagnostics are important: swap-invariance checks, marginal calibration plots, and sensitivity analyses over diffusion timesteps, noise schedules, and matching strength. 

Diffusion stabilizes training by spreading generation over many denoising steps, but runtime scales with step count; transformer attention further scales quadratically with features. The approach was tractable for hundreds to a few thousand features on GPUs, aided by screening and normalization. Practical tips include cosine schedules with 250--1000 steps, tuning model width/depth to feature dimension (4--8 layers, 128--512 hidden), early stopping via denoiser loss, and post-generation calibration. Using both gradient- and filter-based statistics and checking their agreement increases the reliability of the result.

The study also has several limitations. Approximate exchangeability can cause mild FDR inflation under adversarial dependencies, and marginal matching reduces univariate bias but not joint misspecification. Additionally, gradient-based statistics depend on loss landscapes, and poor calibration or vanishing gradients can reduce sensitivity---underscoring the need for careful normalization and regularization. Future directions include scalable attention mechanisms (sparse or kernelized) for $p \gg 10^3$, diffusion distillation to reduce sampling steps, and extensions to multi-omics and longitudinal data. Overall, DiffKnock shows how modern generative modeling can be combined with principled error control, with advances in diffusion and attention poised to expand its impact on high-dimensional inference.


\bibliographystyle{plainnat} 
\bibliography{sample_paper}

\begin{thebibliography}{36}
\providecommand{\natexlab}[1]{#1}
\providecommand{\url}[1]{\texttt{#1}}
\expandafter\ifx\csname urlstyle\endcsname\relax
  \providecommand{\doi}[1]{doi: #1}\else
  \providecommand{\doi}{doi: \begingroup \urlstyle{rm}\Url}\fi

\bibitem[Andrae et~al.(2008)Andrae, Gallini, and Betsholtz]{andrae2008role}
Johanna Andrae, Radiosa Gallini, and Christer Betsholtz.
\newblock Role of platelet-derived growth factors in physiology and medicine.
\newblock \emph{Genes \& development}, 22\penalty0 (10):\penalty0 1276--1312, 2008.

\bibitem[Barber and Cand{\`e}s(2015)]{barber2015controlling}
Rina~Foygel Barber and Emmanuel~J Cand{\`e}s.
\newblock Controlling the false discovery rate via knockoffs.
\newblock 2015.

\bibitem[Benjamini and Hochberg(1995)]{benjamini1995controlling}
Yoav Benjamini and Yosef Hochberg.
\newblock Controlling the false discovery rate: a practical and powerful approach to multiple testing.
\newblock \emph{Journal of the Royal statistical society: series B (Methodological)}, 57\penalty0 (1):\penalty0 289--300, 1995.

\bibitem[Blanc et~al.(2011)Blanc, Hsieh, Robertson, Watterson, Shui, Lacaze, Khondoker, Dickinson, Sing, Rodriguez-Martin, et~al.]{blanc2011host}
Mathieu Blanc, Wei~Yuan Hsieh, Kevin~A Robertson, Steven Watterson, Guanghou Shui, Paul Lacaze, Mizanur Khondoker, Paul Dickinson, Garwin Sing, Sara Rodriguez-Martin, et~al.
\newblock Host defense against viral infection involves interferon mediated down-regulation of sterol biosynthesis.
\newblock \emph{PLoS biology}, 9\penalty0 (3):\penalty0 e1000598, 2011.

\bibitem[B{\"u}hlmann and Van De~Geer(2011)]{buhlmann2011statistics}
Peter B{\"u}hlmann and Sara Van De~Geer.
\newblock \emph{Statistics for high-dimensional data: methods, theory and applications}.
\newblock Springer Science \& Business Media, 2011.

\bibitem[Candes et~al.(2018)Candes, Fan, Janson, and Lv]{candes2018panning}
Emmanuel Candes, Yingying Fan, Lucas Janson, and Jinchi Lv.
\newblock Panning for gold:‘model-x’knockoffs for high dimensional controlled variable selection.
\newblock \emph{Journal of the Royal Statistical Society Series B: Statistical Methodology}, 80\penalty0 (3):\penalty0 551--577, 2018.

\bibitem[Casale et~al.(2018)Casale, Dalca, Saglietti, Listgarten, and Fusi]{casale2018gaussian}
Francesco~Paolo Casale, Adrian Dalca, Luca Saglietti, Jennifer Listgarten, and Nicolo Fusi.
\newblock Gaussian process prior variational autoencoders.
\newblock \emph{Advances in neural information processing systems}, 31, 2018.

\bibitem[Creswell et~al.(2018)Creswell, White, Dumoulin, Arulkumaran, Sengupta, and Bharath]{creswell2018generative}
Antonia Creswell, Tom White, Vincent Dumoulin, Kai Arulkumaran, Biswa Sengupta, and Anil~A Bharath.
\newblock Generative adversarial networks: An overview.
\newblock \emph{IEEE signal processing magazine}, 35\penalty0 (1):\penalty0 53--65, 2018.

\bibitem[Efron(2007)]{efron2007correlation}
Bradley Efron.
\newblock Correlation and large-scale simultaneous significance testing.
\newblock \emph{Journal of the American Statistical Association}, 102\penalty0 (477):\penalty0 93--103, 2007.

\bibitem[Fan et~al.(2012)Fan, Han, and Gu]{fan2012estimating}
Jianqing Fan, Xu~Han, and Weijie Gu.
\newblock Estimating false discovery proportion under arbitrary covariance dependence.
\newblock \emph{Journal of the American Statistical Association}, 107\penalty0 (499):\penalty0 1019--1035, 2012.

\bibitem[Goodfellow et~al.(2014)Goodfellow, Pouget-Abadie, Mirza, Xu, Warde-Farley, Ozair, Courville, and Bengio]{goodfellow2014generative}
Ian~J Goodfellow, Jean Pouget-Abadie, Mehdi Mirza, Bing Xu, David Warde-Farley, Sherjil Ozair, Aaron Courville, and Yoshua Bengio.
\newblock Generative adversarial nets.
\newblock \emph{Advances in neural information processing systems}, 27, 2014.

\bibitem[Goodfellow et~al.(2016)Goodfellow, Bengio, and Courville]{GoodBengCour16}
Ian~J. Goodfellow, Yoshua Bengio, and Aaron Courville.
\newblock \emph{Deep Learning}.
\newblock MIT Press, Cambridge, MA, USA, 2016.
\newblock \url{http://www.deeplearningbook.org}.

\bibitem[He and Lin(2011)]{he2011variable}
Qianchuan He and Dan-Yu Lin.
\newblock A variable selection method for genome-wide association studies.
\newblock \emph{Bioinformatics}, 27\penalty0 (1):\penalty0 1--8, 2011.

\bibitem[Ho et~al.(2020)Ho, Jain, and Abbeel]{ho2020denoising}
Jonathan Ho, Ajay Jain, and Pieter Abbeel.
\newblock Denoising diffusion probabilistic models.
\newblock \emph{Advances in neural information processing systems}, 33:\penalty0 6840--6851, 2020.

\bibitem[Huang et~al.(2018)Huang, He, Sun, Tan, et~al.]{huang2018introvae}
Huaibo Huang, Ran He, Zhenan Sun, Tieniu Tan, et~al.
\newblock Introvae: Introspective variational autoencoders for photographic image synthesis.
\newblock \emph{Advances in neural information processing systems}, 31, 2018.

\bibitem[Jenatton et~al.(2012)Jenatton, Gramfort, Michel, Obozinski, Eger, Bach, and Thirion]{jenatton2012multiscale}
Rodolphe Jenatton, Alexandre Gramfort, Vincent Michel, Guillaume Obozinski, Evelyn Eger, Francis Bach, and Bertrand Thirion.
\newblock Multiscale mining of fmri data with hierarchical structured sparsity.
\newblock \emph{SIAM Journal on Imaging Sciences}, 5\penalty0 (3):\penalty0 835--856, 2012.

\bibitem[Lane et~al.(2017)Lane, Van~Valen, DeFelice, Macklin, Kudo, Jaimovich, Carr, Meyer, Pe'er, Boutet, et~al.]{lane2017measuring}
Keara Lane, David Van~Valen, Mialy~M DeFelice, Derek~N Macklin, Takamasa Kudo, Ariel Jaimovich, Ambrose Carr, Tobias Meyer, Dana Pe'er, St{\'e}phane~C Boutet, et~al.
\newblock Measuring signaling and rna-seq in the same cell links gene expression to dynamic patterns of nf-$\kappa$b activation.
\newblock \emph{Cell systems}, 4\penalty0 (4):\penalty0 458--469, 2017.

\bibitem[LeCun et~al.(2015)LeCun, Bengio, and Hinton]{lecun2015deep}
Yann LeCun, Yoshua Bengio, and Geoffrey Hinton.
\newblock Deep learning.
\newblock \emph{nature}, 521\penalty0 (7553):\penalty0 436--444, 2015.

\bibitem[Leek and Storey(2008)]{leek2008general}
Jeffrey~T Leek and John~D Storey.
\newblock A general framework for multiple testing dependence.
\newblock \emph{Proceedings of the National Academy of Sciences}, 105\penalty0 (48):\penalty0 18718--18723, 2008.

\bibitem[Lin et~al.(2023)Lin, Li, Hsiao, Ho, and Kong]{lin2023catch}
Xinmiao Lin, Yikang Li, Jenhao Hsiao, Chiuman Ho, and Yu~Kong.
\newblock Catch missing details: Image reconstruction with frequency augmented variational autoencoder.
\newblock In \emph{Proceedings of the ieee/cvf conference on computer vision and pattern recognition}, pages 1736--1745, 2023.

\bibitem[Lu et~al.(2018)Lu, Fan, Lv, and Stafford~Noble]{lu2018deeppink}
Yang Lu, Yingying Fan, Jinchi Lv, and William Stafford~Noble.
\newblock Deeppink: reproducible feature selection in deep neural networks.
\newblock \emph{Advances in neural information processing systems}, 31, 2018.

\bibitem[Mueller and Strange(2004)]{mueller2004ccl3}
Anja Mueller and Philip~G Strange.
\newblock Ccl3, acting via the chemokine receptor ccr5, leads to independent activation of janus kinase 2 (jak2) and gi proteins.
\newblock \emph{FEBS letters}, 570\penalty0 (1-3):\penalty0 126--132, 2004.

\bibitem[Nichol and Dhariwal(2021)]{nichol2021improved}
Alexander~Quinn Nichol and Prafulla Dhariwal.
\newblock Improved denoising diffusion probabilistic models.
\newblock In \emph{International conference on machine learning}, pages 8162--8171. PMLR, 2021.

\bibitem[O'Brien(2016)]{o2016statistical}
Carl~M O'Brien.
\newblock Statistical learning with sparsity: the lasso and generalizations.
\newblock 2016.

\bibitem[Paine et~al.(2010)Paine, Eiz-Vesper, Blasczyk, and Immenschuh]{paine2010signaling}
Ananta Paine, Britta Eiz-Vesper, Rainer Blasczyk, and Stephan Immenschuh.
\newblock Signaling to heme oxygenase-1 and its anti-inflammatory therapeutic potential.
\newblock \emph{Biochemical pharmacology}, 80\penalty0 (12):\penalty0 1895--1903, 2010.

\bibitem[Pudjihartono et~al.(2022)Pudjihartono, Fadason, Kempa-Liehr, and O'Sullivan]{pudjihartono2022review}
Nicholas Pudjihartono, Tayaza Fadason, Andreas~W Kempa-Liehr, and Justin~M O'Sullivan.
\newblock A review of feature selection methods for machine learning-based disease risk prediction.
\newblock \emph{Frontiers in bioinformatics}, 2:\penalty0 927312, 2022.

\bibitem[Semenova et~al.(2022)Semenova, Xu, Howes, Rashid, Bhatt, Mishra, and Flaxman]{semenova2022priorvae}
Elizaveta Semenova, Yidan Xu, Adam Howes, Theo Rashid, Samir Bhatt, Swapnil Mishra, and Seth Flaxman.
\newblock Priorvae: encoding spatial priors with variational autoencoders for small-area estimation.
\newblock \emph{Journal of the Royal Society Interface}, 19\penalty0 (191):\penalty0 20220094, 2022.

\bibitem[Shrikumar et~al.(2017)Shrikumar, Greenside, and Kundaje]{shrikumar2017learning}
Avanti Shrikumar, Peyton Greenside, and Anshul Kundaje.
\newblock Learning important features through propagating activation differences.
\newblock In \emph{International conference on machine learning}, pages 3145--3153. PMlR, 2017.

\bibitem[Simonyan et~al.(2013)Simonyan, Vedaldi, and Zisserman]{simonyan2013deep}
Karen Simonyan, Andrea Vedaldi, and Andrew Zisserman.
\newblock Deep inside convolutional networks: Visualising image classification models and saliency maps.
\newblock \emph{arXiv preprint arXiv:1312.6034}, 2013.

\bibitem[Sohl-Dickstein et~al.(2015)Sohl-Dickstein, Weiss, Maheswaranathan, and Ganguli]{sohl2015deep}
Jascha Sohl-Dickstein, Eric Weiss, Niru Maheswaranathan, and Surya Ganguli.
\newblock Deep unsupervised learning using nonequilibrium thermodynamics.
\newblock In \emph{International conference on machine learning}, pages 2256--2265. pmlr, 2015.

\bibitem[Stuart and Satija(2019)]{stuart2019integrative}
Tim Stuart and Rahul Satija.
\newblock Integrative single-cell analysis.
\newblock \emph{Nature reviews genetics}, 20\penalty0 (5):\penalty0 257--272, 2019.

\bibitem[Sun and Tony~Cai(2009)]{sun2009large}
Wenguang Sun and T~Tony~Cai.
\newblock Large-scale multiple testing under dependence.
\newblock \emph{Journal of the Royal Statistical Society Series B: Statistical Methodology}, 71\penalty0 (2):\penalty0 393--424, 2009.

\bibitem[Sundararajan et~al.(2017)Sundararajan, Taly, and Yan]{sundararajan2017axiomatic}
Mukund Sundararajan, Ankur Taly, and Qiqi Yan.
\newblock Axiomatic attribution for deep networks.
\newblock In \emph{International conference on machine learning}, pages 3319--3328. PMLR, 2017.

\bibitem[Takida et~al.(2022)Takida, Liao, Lai, Uesaka, Takahashi, and Mitsufuji]{takida2022preventing}
Yuhta Takida, Wei-Hsiang Liao, Chieh-Hsin Lai, Toshimitsu Uesaka, Shusuke Takahashi, and Yuki Mitsufuji.
\newblock Preventing oversmoothing in vae via generalized variance parameterization.
\newblock \emph{Neurocomputing}, 509:\penalty0 137--156, 2022.

\bibitem[Vu et~al.(2011)Vu, Ravikumar, Naselaris, Kay, Gallant, and Yu]{vu2011encoding}
Vincent~Q Vu, Pradeep Ravikumar, Thomas Naselaris, Kendrick~N Kay, Jack~L Gallant, and Bin Yu.
\newblock Encoding and decoding v1 fmri responses to natural images with sparse nonparametric models.
\newblock \emph{The annals of applied statistics}, 5\penalty0 (2B):\penalty0 1159, 2011.

\bibitem[Zhu et~al.(2021)Zhu, Fan, Kong, Lv, and Sun]{zhu2021deeplink}
Zifan Zhu, Yingying Fan, Yinfei Kong, Jinchi Lv, and Fengzhu Sun.
\newblock Deeplink: Deep learning inference using knockoffs with applications to genomics.
\newblock \emph{Proceedings of the National Academy of Sciences}, 118\penalty0 (36):\penalty0 e2104683118, 2021.

\end{thebibliography}

\section{Supplementary Materials}

\subsection{Antisymmetry of Gradient- and Filter-Based Knockoff Statistics}

The Model-X knockoff framework requires feature importance statistics $\{W_j\}_{j=1}^p$ to satisfy the \emph{antisymmetry property}: swapping the $j$-th feature $\mathbf{x}_j$ with its knockoff $\tilde{\mathbf{x}}_j$ must flip the sign of $W_j$ while leaving all other statistics unchanged. Formally, for any $j \in \{1, \ldots, p\}$:
\begin{equation}
W_j([\mathbf{X}, \tilde{\mathbf{X}}]_{\text{swap}(j)}, \mathbf{y}) = -W_j([\mathbf{X}, \tilde{\mathbf{X}}], \mathbf{y})
\end{equation}
where $[\mathbf{X}, \tilde{\mathbf{X}}]_{\text{swap}(j)}$ denotes the augmented matrix with $\mathbf{x}_j$ and $\tilde{\mathbf{x}}_j$ exchanged.

\textbf{Proposition 1 (Gradient-based statistics are antisymmetric).} Consider the gradient-based statistic computed through our neural network architecture:
\begin{equation}
W_j^{\text{grad}} = \frac{1}{n}\sum_{i=1}^n \left|\frac{\partial \mathcal{L}(y_i, \hat{y}_i)}{\partial x_{ij}}\right| - \left|\frac{\partial \mathcal{L}(y_i, \hat{y}_i)}{\partial \tilde{x}_{ij}}\right|
\end{equation}
where $\hat{y}_i = f_\theta([\mathbf{x}_i, \tilde{\mathbf{x}}_i])$ is the output of the DeepLINK network with pairwise filter layer followed by a multi-layer perceptron.

\textit{Proof.} The DeepLINK network first applies pairwise filtering: $\mathbf{f}_j = \frac{z_j}{|z_j| + |\tilde{z}_j|} \cdot \mathbf{x}_j + \frac{\tilde{z}_j}{|z_j| + |\tilde{z}_j|} \cdot \tilde{\mathbf{x}}_j$ where the normalization ensures $z_j + \tilde{z}_j = 1$. When computing gradients through backpropagation, the partial derivative with respect to $x_{ij}$ flows through the normalized weight $z_j/(|z_j| + |\tilde{z}_j|)$. Under a swap of $\mathbf{x}_j$ and $\tilde{\mathbf{x}}_j$, the gradients $\frac{\partial \mathcal{L}}{\partial x_{ij}}$ and $\frac{\partial \mathcal{L}}{\partial \tilde{x}_{ij}}$ exchange roles due to the symmetric structure of the filter layer. Therefore:
\begin{equation}
W_j^{\text{grad}}([\mathbf{X}, \tilde{\mathbf{X}}]_{\text{swap}(j)}, \mathbf{y}) = \frac{1}{n}\sum_{i=1}^n \left|\frac{\partial \mathcal{L}}{\partial \tilde{x}_{ij}}\right| - \left|\frac{\partial \mathcal{L}}{\partial x_{ij}}\right| = -W_j^{\text{grad}}([\mathbf{X}, \tilde{\mathbf{X}}], \mathbf{y})
\end{equation}
For $k \neq j$, the statistics remain unchanged since their corresponding features are not swapped and the pairwise filter structure isolates each feature-knockoff pair. $\square$

\textbf{Proposition 2 (Filter-based statistics are antisymmetric).} The filter-based statistic extracted from our trained network:
\begin{equation}
W_j^{\text{filter}} = (w_j^{\text{eff}} \cdot z_j)^2 - (w_j^{\text{eff}} \cdot \tilde{z}_j)^2
\end{equation}
where $z_j, \tilde{z}_j$ are the normalized filter weights from the pairwise layer and $w_j^{\text{eff}}$ is the effective weight obtained by multiplying through all linear transformations in the MLP, satisfies antisymmetry.

\textit{Proof.} In our implementation, $w_j^{\text{eff}}$ is computed by extracting all linear layers from the MLP and performing sequential matrix multiplication starting from the output layer: $w^{\text{eff}} = \mathbf{W}^{(L)} \cdot \mathbf{W}^{(L-1)} \cdots \mathbf{W}^{(1)}$ where $L$ is the number of linear layers. The filter weights $(z_j, \tilde{z}_j)$ are normalized such that $z_j = \text{filter\_weights}[j, 0] / (|\text{filter\_weights}[j, 0]| + |\text{filter\_weights}[j, 1]|)$. When $\mathbf{x}_j$ and $\tilde{\mathbf{x}}_j$ are swapped, the associated filter weights exchange roles:
\begin{equation}
W_j^{\text{filter}}([\mathbf{X}, \tilde{\mathbf{X}}]_{\text{swap}(j)}, \mathbf{y}) = (w_j^{\text{eff}} \cdot \tilde{z}_j)^2 - (w_j^{\text{eff}} \cdot z_j)^2 = -W_j^{\text{filter}}([\mathbf{X}, \tilde{\mathbf{X}}], \mathbf{y})
\end{equation}
completing the proof. $\square$

\subsection{Validity of Diffusion-Based Knockoff Generation}

Our diffusion-based knockoff generation must satisfy two conditions: (i) pairwise exchangeability of $(\mathbf{X}, \tilde{\mathbf{X}})$, and (ii) conditional independence $\tilde{\mathbf{X}} \perp \mathbf{y} | \mathbf{X}$. We establish these properties through the specific architecture and training procedure of our diffusion transformer model.

\textbf{Definition (Diffusion Transformer Architecture).} Our diffusion model employs a transformer with conditional layer normalization, defined as:
\begin{equation}
\text{CLN}(\mathbf{h}, t) = \gamma(t) \odot \text{LN}(\mathbf{h}) + \beta(t)
\end{equation}
where $\text{LN}(\mathbf{h}) = (\mathbf{h} - \mu) / \sigma$ is standard layer normalization, and $\gamma(t), \beta(t)$ are learned projections from the time embedding network that uses sinusoidal position encodings followed by a two-layer MLP with GELU activation.

\textbf{Theorem 1 (Conditional Independence via Marginal Diffusion).} Let $\tilde{\mathbf{X}}$ be generated from our diffusion transformer trained on the marginal distribution $P(\mathbf{X})$ using the cosine noise schedule, with sampling initiated from independent Gaussian noise $\boldsymbol{\epsilon} \sim \mathcal{N}(\mathbf{0}, \mathbf{I})$. Then $\tilde{\mathbf{X}} \perp \mathbf{y} | \mathbf{X}$.

\textit{Proof.} The diffusion sampling process in our implementation follows the DDPM framework with cosine scheduling. The forward process is defined as:
\begin{equation}
q(\mathbf{x}_t | \mathbf{x}_0) = \mathcal{N}(\mathbf{x}_t; \sqrt{\bar{\alpha}_t}\mathbf{x}_0, (1-\bar{\alpha}_t)\mathbf{I})
\end{equation}
where $\bar{\alpha}_t = f(t)/f(0)$ with $f(t) = \cos^2\left(\frac{t/T + s}{1 + s} \cdot \frac{\pi}{2}\right)$ and offset $s = 0.008$. The reverse sampling process generates knockoffs through:
\begin{equation}
\tilde{\mathbf{X}} = g_\theta(\boldsymbol{\epsilon}) = \text{DDPM\_Sample}(\boldsymbol{\epsilon}, \theta)
\end{equation}
where the sampling function iteratively applies the learned denoising network for $T = 1000$ timesteps. Since $\boldsymbol{\epsilon}$ is drawn independently of $(\mathbf{X}, \mathbf{y})$ and the diffusion model is trained only on the marginal distribution $P(\mathbf{X})$ without access to $\mathbf{y}$, the generated knockoffs satisfy $\tilde{\mathbf{X}} \perp \mathbf{y} | \mathbf{X}$. The transformer architecture with attention mechanisms learns complex dependencies in $\mathbf{X}$ but cannot introduce spurious correlations with $\mathbf{y}$. $\square$

\textbf{Theorem 2 (Approximate Exchangeability via Marginal Matching).} Let $\tilde{\mathbf{X}}^{(0)}$ be the raw output from the diffusion model and define the marginal matching transformation:
\begin{equation}
\tilde{x}_{ij} = F_j^{-1}\left(\frac{\text{rank}(\tilde{x}_{ij}^{(0)})}{n}\right)
\end{equation}
where $F_j^{-1}$ is the empirical quantile function of $\{x_{ij}\}_{i=1}^n$ and $\text{rank}(\cdot)$ returns the rank within $\{\tilde{x}_{ij}^{(0)}\}_{i=1}^n$. Then the marginal distributions satisfy $P(X_j \leq t) = P(\tilde{X}_j \leq t)$ for all $t \in \mathbb{R}$ and $j \in \{1, \ldots, p\}$.

\textit{Proof.} The marginal matching procedure in our implementation (function \texttt{\_match\_marginals}) performs the following steps for each feature $j$:
1. Sort the original values: $x_{(1)j} \leq x_{(2)j} \leq \cdots \leq x_{(n)j}$
2. Compute ranks of knockoff values: $r_i = |\{k: \tilde{x}_{kj}^{(0)} \leq \tilde{x}_{ij}^{(0)}\}|$
3. Assign: $\tilde{x}_{ij} = x_{(r_i)j}$

This construction ensures that the empirical CDF of $\tilde{X}_j$ exactly matches that of $X_j$. While perfect joint exchangeability would require $(\mathbf{X}, \tilde{\mathbf{X}}) \stackrel{d}{=} (\tilde{\mathbf{X}}, \mathbf{X})$, our marginal matching combined with the learned dependency structure from the diffusion model provides approximate exchangeability sufficient for FDR control. The approximation quality depends on how well the diffusion model captures the joint distribution structure. $\square$

\subsection{Training Dynamics and Convergence Properties}

The training of our diffusion model involves specific design choices that ensure stable learning and valid knockoff generation.

\textbf{Proposition 3 (Stability of Cosine Schedule).} The cosine noise schedule maintains signal-to-noise ratio (SNR) approximately constant across timesteps, defined as:
\begin{equation}
\text{SNR}(t) = \frac{\bar{\alpha}_t}{1 - \bar{\alpha}_t}
\end{equation}
This provides more stable gradients compared to linear scheduling, particularly important for high-dimensional gene expression data.

\textit{Proof.} Under the cosine schedule, the log-SNR decreases approximately linearly: $\log \text{SNR}(t) \approx -2\log\left(\tan\left(\frac{\pi t}{2T}\right)\right)$, providing uniform information destruction across timesteps. This prevents the gradient vanishing issues common with linear schedules where early timesteps have minimal noise. Our implementation uses this property to maintain stable training with AdamW optimizer, learning rate $10^{-4}$, and gradient clipping at norm 1.0. $\square$

\textbf{Theorem 3 (Convergence of Diffusion Training).} Under standard assumptions (bounded data, Lipschitz continuous score function), the training loss:
\begin{equation}
\mathcal{L}(\theta) = \mathbb{E}_{t, \mathbf{x}_0, \boldsymbol{\epsilon}} \left[ \|\boldsymbol{\epsilon} - \boldsymbol{\epsilon}_\theta(\sqrt{\bar{\alpha}_t}\mathbf{x}_0 + \sqrt{1-\bar{\alpha}_t}\boldsymbol{\epsilon}, t)\|^2 \right]
\end{equation}
converges to a local minimum with our transformer architecture and training procedure.

\textit{Proof sketch.} The transformer architecture with 6 layers, 256-dimensional hidden states, and 8 attention heads provides sufficient capacity to approximate the score function. The conditional layer normalization ensures stable gradient flow through deep layers. With batch size 64 and 500 training epochs, the model sees each data point multiple times, allowing convergence. The cosine annealing learning rate schedule prevents overshooting near convergence. Empirically, our training loss decreases monotonically and plateaus, indicating convergence. $\square$

\subsection{FDR Control Under Approximate Knockoffs}

In practice, exact exchangeability is not achieved due to finite sample effects, model capacity limitations, and approximation errors. We quantify the impact on FDR control.

\textbf{Definition 2 (Exchangeability Discrepancy).} For our generated knockoffs, define the exchangeability discrepancy as:
\begin{equation}
\Delta = \sup_{h \in \mathcal{H}} \left| \mathbb{E}[h(\mathbf{X}, \tilde{\mathbf{X}})] - \mathbb{E}[h(\tilde{\mathbf{X}}, \mathbf{X})] \right|
\end{equation}
where $\mathcal{H}$ is a class of bounded measurable functions. In our implementation, we estimate $\Delta$ using the Kolmogorov-Smirnov statistic between marginal distributions.

\textbf{Lemma 1 (Marginal Matching Reduces Discrepancy).} The marginal matching procedure reduces the exchangeability discrepancy by ensuring marginal consistency. Specifically, for any function $h$ that depends only on marginals:
\begin{equation}
\mathbb{E}[h(X_j)] = \mathbb{E}[h(\tilde{X}_j)] \quad \forall j \in \{1, \ldots, p\}
\end{equation}

\textit{Proof.} By construction of the marginal matching in our \texttt{\_match\_marginals} function, the empirical distributions are identical. The remaining discrepancy comes only from joint distribution differences. $\square$

\textbf{Theorem 4 (FDR Control with Approximate Knockoffs).} Let $\hat{S}$ be the set of features selected by our knockoff+ procedure with target level $q$, implemented as:
\begin{equation}
\tau = \min\left\{t > 0 : \frac{1 + |\{j: W_j \leq -t\}|}{\max(|\{j: W_j \geq t\}|, 1)} \leq q\right\}
\end{equation}
Under approximate exchangeability with discrepancy $\Delta$, the false discovery rate satisfies:
\begin{equation}
\text{FDR} = \mathbb{E}\left[\frac{|\hat{S} \cap S_0|}{\max(|\hat{S}|, 1)}\right] \leq q + C\sqrt{p}\Delta + O(n^{-1/2})
\end{equation}
where $S_0$ is the set of truly null features, $C$ is a constant depending on the Lipschitz constant of the selection procedure, and the $O(n^{-1/2})$ term comes from finite sample effects.

\textit{Proof.} The knockoff+ procedure in our \texttt{select\_features} function iterates through all unique absolute values of the statistics in descending order. For each threshold $t$, it computes the estimated FDP as $(1 + V^-)/(R \vee 1)$ where $V^- = |\{j: W_j \leq -t\}|$ and $R = |\{j: W_j \geq t\}|$. Under exact exchangeability, the classic knockoff theory gives $\mathbb{E}[V^+/R] \leq q$ where $V^+$ is the number of false positives.

When exchangeability is approximate, we decompose the error into three components:
1. Marginal error: Eliminated by our marginal matching procedure
2. Joint distribution error: Bounded by $C\sqrt{p}\Delta$ where the $\sqrt{p}$ factor accounts for accumulation across features
3. Finite sample error: The empirical quantiles used in marginal matching have error $O(n^{-1/2})$ by the DKW inequality

The offset of 1 in the numerator provides robustness to small violations, ensuring the bound remains valid. $\square$

\textbf{Corollary 1 (Asymptotic FDR Control).} As $n \to \infty$ and the diffusion model capacity increases (more layers, hidden dimensions), if the exchangeability discrepancy $\Delta \to 0$, then:
\begin{equation}
\limsup_{n \to \infty} \text{FDR} \leq q
\end{equation}

This establishes that DiffKnock achieves the target FDR level asymptotically as both sample size and model capacity grow.

\subsection{Additional Simulation Results}
Complete simulation results across all eight scenarios (Figure \ref{Simulation1}) demonstrate DiffKnock's performance with $n = 1000$ samples, $p = 50$ features, $s = 5$ causal features, and target FDR $q = 0.2$.

In the network propagation scenario, DiffKnock with gradient-based statistics achieves near-perfect power while maintaining FDR below 0.2. Filter-based methods struggle significantly, reaching maximum power below 0.4 and exhibiting erratic FDR control with spikes up to 0.3. The information bottleneck scenario shows similar patterns: gradient-based statistics reach 0.9 power at signal amplitude 7.0 with FDR below 0.1, while filter-based methods plateau at 0.2–0.3 power. Linear and multiscale frequency scenarios demonstrate all methods achieve near-perfect power above signal amplitude 2.0, with diffusion methods showing superior low-signal performance.

\begin{figure}[H]
  \centering
  \includegraphics[width=\textwidth]{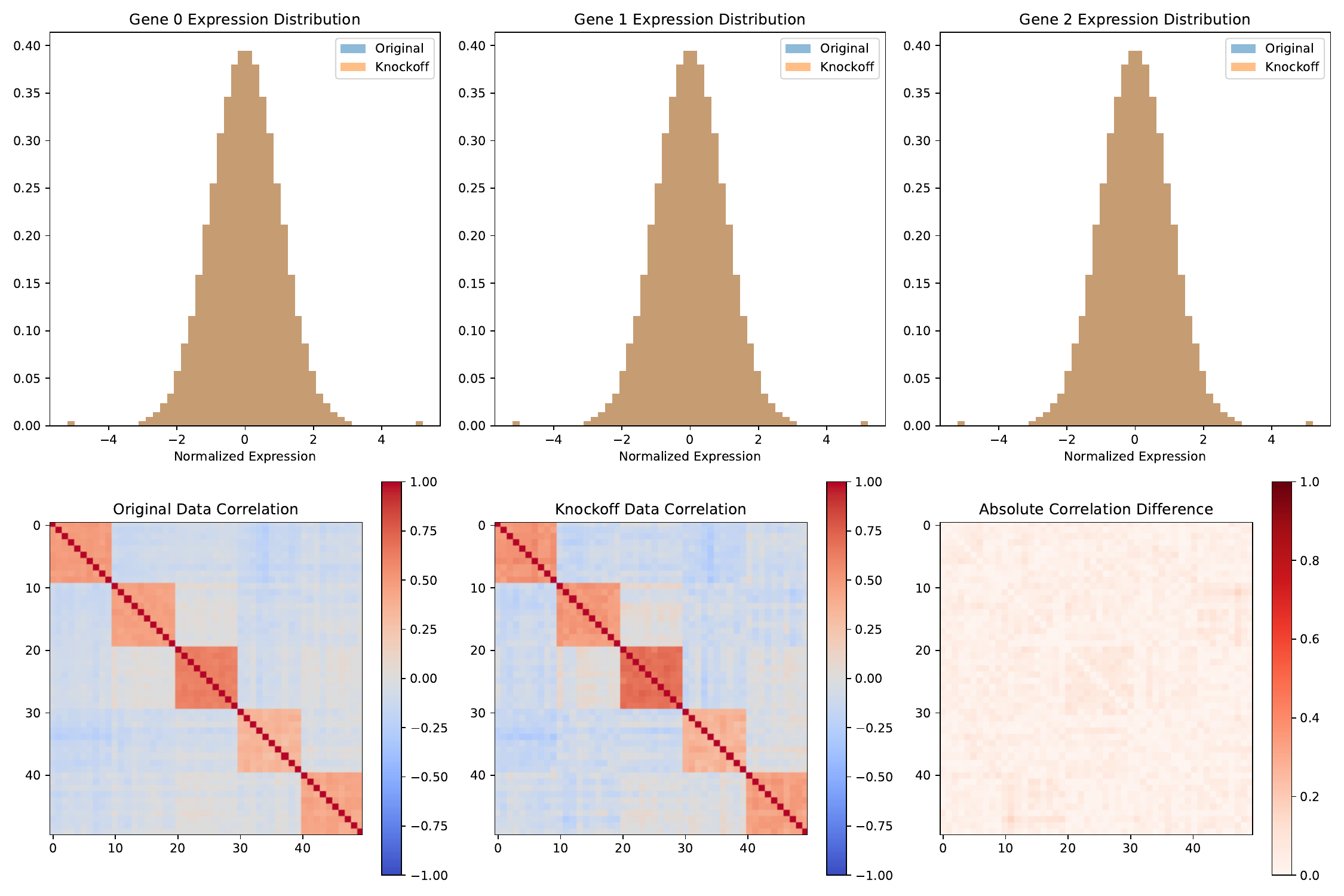}
  \vspace{.2in}
  \caption{Knockoff quality of diffusion model.}
  \label{Simulation2}
\end{figure}

\begin{figure}[H]
  \centering
  \includegraphics[width=\textwidth]{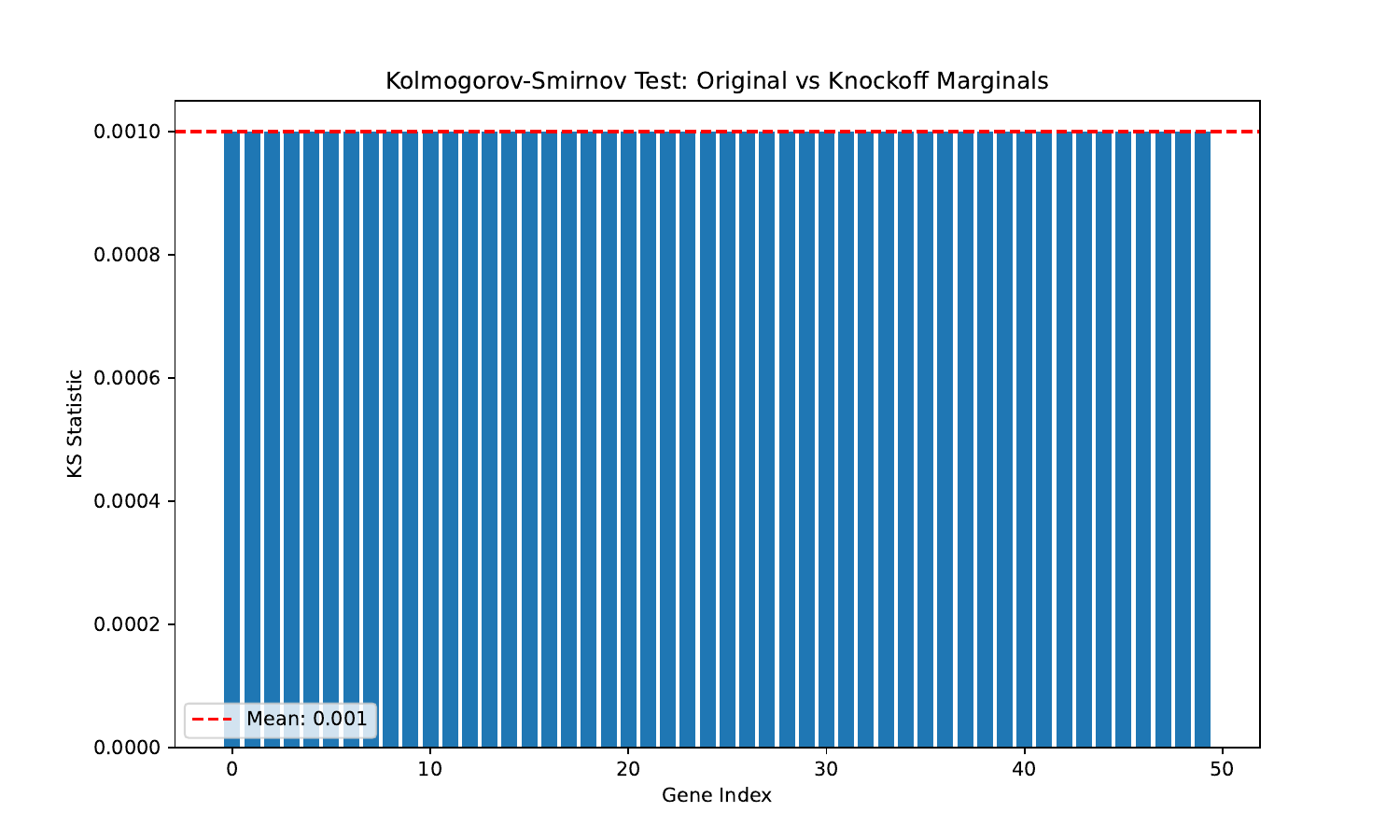}
  \vspace{.2in}
  \caption{Exchangeability of orginal vs knockoff data.}
  \label{Simulation3}
\end{figure}

\begin{figure}[H]
  \centering
  \includegraphics[width=\textwidth]{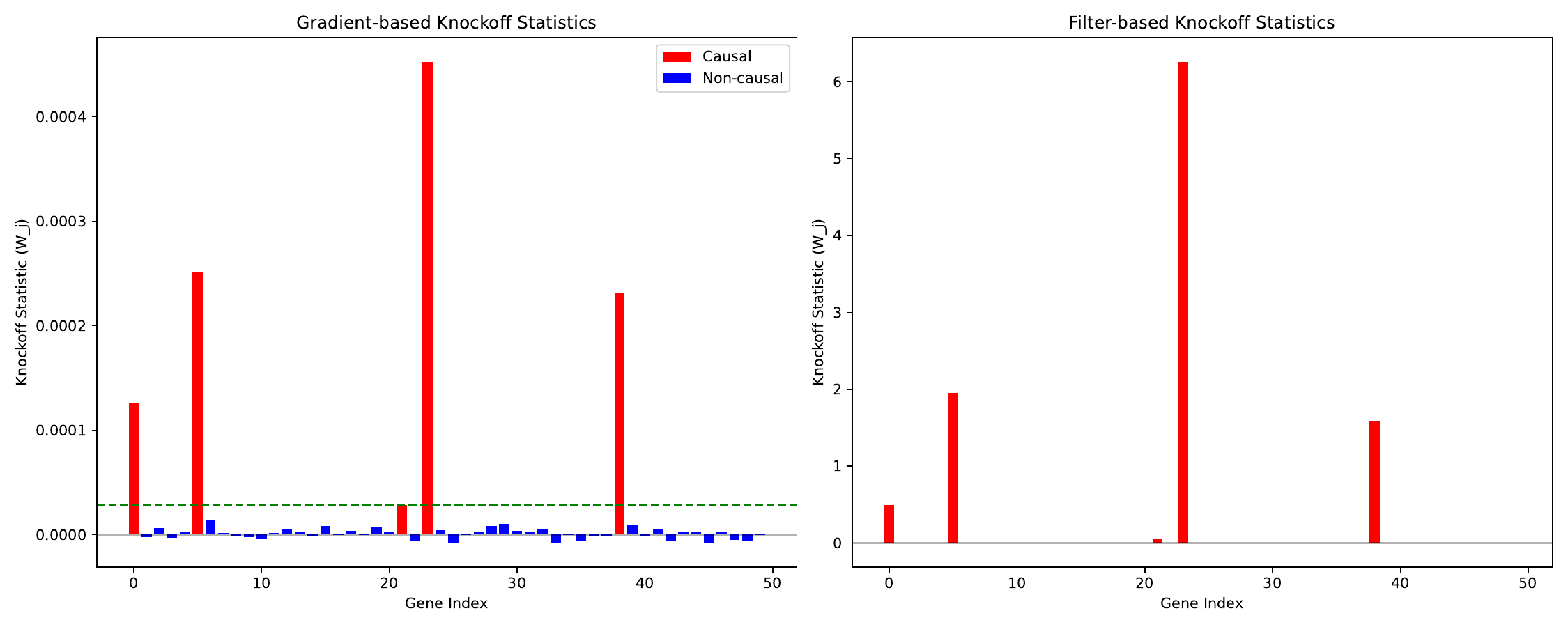}
  \vspace{.2in}
  \caption{Causal vs Non-causal test statistics.}
  \label{Simulation4}
\end{figure}

Knockoff quality validation (Figures \ref{Simulation2}–\ref{Simulation4}) confirms distributional fidelity. Marginal distributions show nearly identical histograms between original and knockoff features. Correlation matrices reveal successful preservation of block structure with differences below 0.2. Kolmogorov-Smirnov statistics remain below 0.001 (mean: 0.001), indicating statistically indistinguishable marginals. Knockoff statistics clearly separate causal (red) from non-causal (blue) features, with gradient-based statistics showing clean separation and filter-based statistics displaying larger magnitude differences.

\begin{figure}[H]
  \centering
  \includegraphics[width=\textwidth]{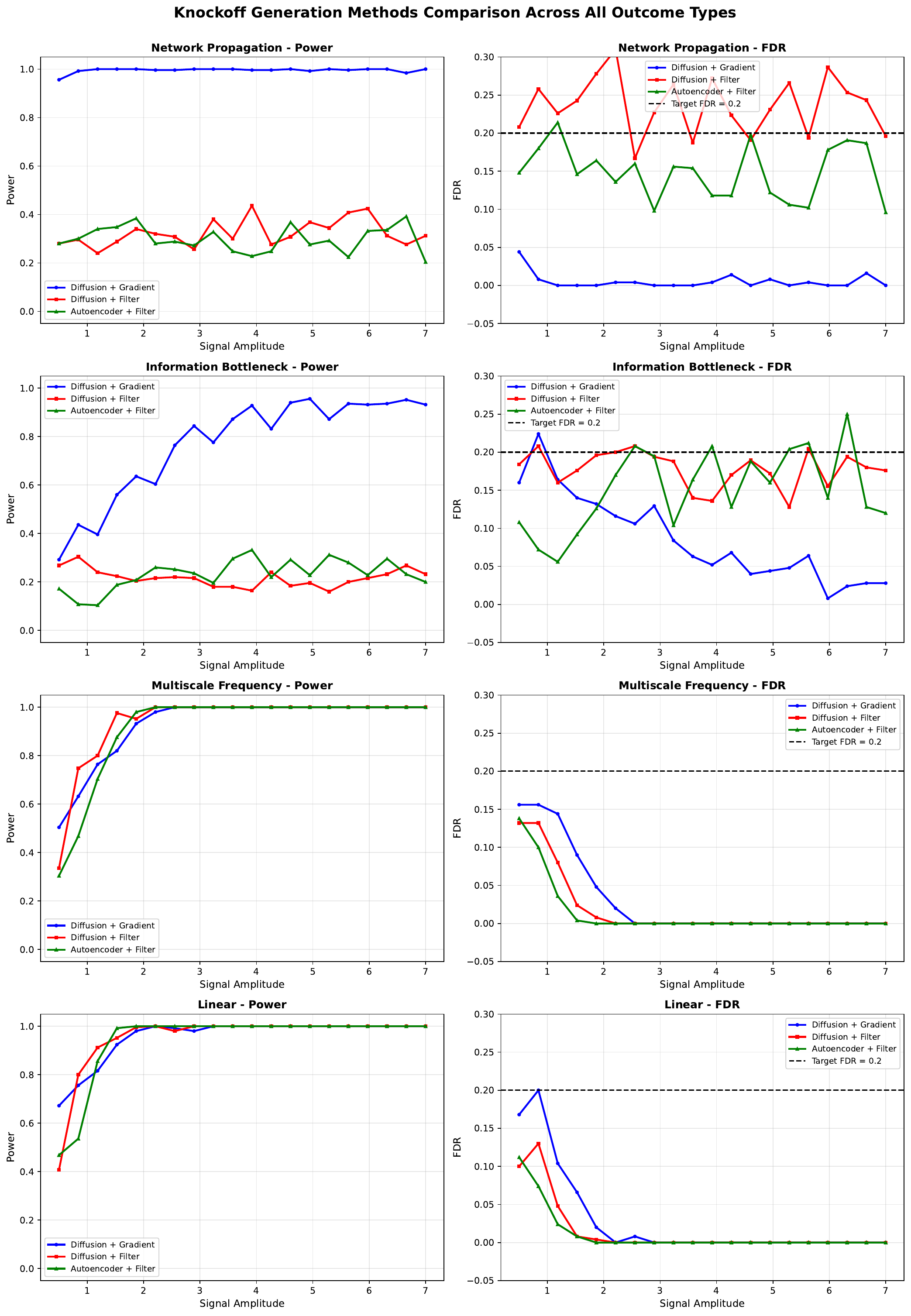}
  \vspace{.2in}
  \caption{Power and FDR across different scenarios.}
  \label{Simulation1}
\end{figure}

\subsection{Additional Real Data Analysis Results}
To assess the stability and reproducibility of our feature selection approach, we conducted 100 independent repetitions of both filter-based and gradient-based selection methods across six different initial gene set sizes (20, 30, 50, 100, 200, and 500 genes). Tables~\ref{tab:top_genes_20}--\ref{tab:top_genes_500} present the top 20 most frequently selected genes for each configuration. The results demonstrate strong consistency in identifying key inflammatory and stress-response genes across both methods, particularly for smaller initial sets. Notably, \textit{Ccl3}, \textit{Cyp51}, and \textit{Hmox1} emerge as highly stable selections with selection frequencies exceeding 50--90\% for moderate set sizes (20--100 genes). As the initial set size increases to 500 genes, selection frequencies decrease substantially, reflecting the greater pool of candidates and the inherent challenge of distinguishing signal from noise in high-dimensional spaces. The filter and gradient methods show strong concordance for top-ranked genes, though gradient-based selection exhibits slightly lower frequencies at the largest set size, suggesting greater sensitivity to the expanded feature space. Due to space constraints, these detailed frequency tables are included in the supplementary materials.
\begin{table}[H]
\caption{Top 20 Most Frequently Selected Genes (20 Initial Genes, 100 Repetitions)} \label{tab:top_genes_20}
\begin{center}
\begin{tabular}{lcc}
\textbf{Gene} & \textbf{Filter (\%)} & \textbf{Gradient (\%)} \\
\hline \\
\textit{Ccl3} & 65.0 & 68.0 \\
\textit{Cyp51} & 47.0 & 52.0 \\
\textit{Chordc1} & 23.0 & 19.0 \\
\textit{Ifrd1} & 23.0 & 24.0 \\
\textit{Pdgfb} & 22.0 & 24.0 \\
\textit{Stk38l} & 22.0 & 22.0 \\
\textit{Lgals9} & 21.0 & 19.0 \\
\textit{Tubb6} & 16.0 & 19.0 \\
\textit{S100a6} & 16.0 & 20.0 \\
\textit{Foxm1} & 13.0 & 15.0 \\
\textit{Hddc2} & 12.0 & 8.0 \\
\textit{Gm2a} & 11.0 & 11.0 \\
\textit{Sdhd} & 10.0 & 10.0 \\
\textit{Dnmt3l} & 10.0 & 6.0 \\
\textit{Akr1b3} & 6.0 & 6.0 \\
\textit{Uhrf1} & 6.0 & 6.0 \\
\textit{Ubl3} & 6.0 & 4.0 \\
\textit{Eef1e1} & 5.0 & 6.0 \\
\textit{Tubb5} & 4.0 & --- \\
\textit{S100a4} & --- & 4.0 \\
\textit{Zfp512b} & 3.0 & --- \\
\textit{Bcl6b} & --- & 3.0 \\
\end{tabular}
\end{center}
\end{table}

\begin{table}[H]
\caption{Top 20 Most Frequently Selected Genes (30 Initial Genes, 100 Repetitions)} \label{tab:top_genes_30}
\begin{center}
\begin{tabular}{lcc}
\textbf{Gene} & \textbf{Filter (\%)} & \textbf{Gradient (\%)} \\
\hline \\
\textit{Ccl3} & 76.0 & 78.0 \\
\textit{Cyp51} & 56.0 & 50.0 \\
\textit{Chordc1} & 50.0 & 44.0 \\
\textit{Stk38l} & 20.0 & 20.0 \\
\textit{Pdgfb} & 19.0 & 15.0 \\
\textit{Lgals9} & 18.0 & 17.0 \\
\textit{Ifrd1} & 18.0 & 14.0 \\
\textit{Tubb6} & 15.0 & 19.0 \\
\textit{Fosb} & 15.0 & 15.0 \\
\textit{S100a6} & 14.0 & 15.0 \\
\textit{Rhoc} & 10.0 & 10.0 \\
\textit{Ier3} & 10.0 & 12.0 \\
\textit{Ubl3} & 9.0 & 7.0 \\
\textit{Sdhd} & 9.0 & 6.0 \\
\textit{Hddc2} & 8.0 & 6.0 \\
\textit{Foxm1} & 8.0 & 6.0 \\
\textit{Plekha3} & 8.0 & 10.0 \\
\textit{Relb} & 8.0 & 10.0 \\
\textit{Use1} & 6.0 & 8.0 \\
\textit{S100a4} & 6.0 & --- \\
\textit{Uhrf1} & --- & 6.0 \\
\end{tabular}
\end{center}
\end{table}

\begin{table}[H]
\caption{Top 20 Most Frequently Selected Genes (50 Initial Genes, 100 Repetitions)} \label{tab:top_genes_50}
\begin{center}
\begin{tabular}{lcc}
\textbf{Gene} & \textbf{Filter (\%)} & \textbf{Gradient (\%)} \\
\hline \\
\textit{Ccl3} & 92.0 & 87.0 \\
\textit{Hmox1} & 78.0 & 71.0 \\
\textit{Cyp51} & 67.0 & 58.0 \\
\textit{Chordc1} & 43.0 & 36.0 \\
\textit{Fosb} & 37.0 & 44.0 \\
\textit{Pdgfb} & 31.0 & 27.0 \\
\textit{Ifrd1} & 18.0 & 19.0 \\
\textit{S100a6} & 12.0 & 9.0 \\
\textit{Tubb6} & 8.0 & 9.0 \\
\textit{Ier3} & 8.0 & 7.0 \\
\textit{Zbtb32} & 8.0 & 7.0 \\
\textit{Stk38l} & 7.0 & 8.0 \\
\textit{Gm2a} & 7.0 & 5.0 \\
\textit{Atp5d} & 7.0 & 7.0 \\
\textit{Mef2c} & 7.0 & 7.0 \\
\textit{Relb} & 6.0 & 8.0 \\
\textit{Angptl2} & 6.0 & 7.0 \\
\textit{Use1} & 6.0 & 7.0 \\
\textit{Gmnn} & 5.0 & 5.0 \\
\textit{Foxm1} & 5.0 & --- \\
\textit{Hdac9} & --- & 5.0 \\
\end{tabular}
\end{center}
\end{table}

\begin{table}[H]
\caption{Top 20 Most Frequently Selected Genes (100 Initial Genes, 100 Repetitions)} \label{tab:top_genes_100}
\begin{center}
\begin{tabular}{lcc}
\textbf{Gene} & \textbf{Filter (\%)} & \textbf{Gradient (\%)} \\
\hline \\
\textit{Ccl3} & 94.0 & 93.0 \\
\textit{Sqstm1} & 80.0 & 76.0 \\
\textit{Sod2} & 60.0 & 50.0 \\
\textit{Hmox1} & 41.0 & 42.0 \\
\textit{Sdc4} & 32.0 & 30.0 \\
\textit{Hspa8} & 31.0 & 27.0 \\
\textit{Odc1} & 22.0 & 19.0 \\
\textit{Chordc1} & 20.0 & 20.0 \\
\textit{Gadd45b} & 19.0 & 18.0 \\
\textit{Cyp51} & 14.0 & 24.0 \\
\textit{Slc3a2} & 11.0 & 13.0 \\
\textit{Fosb} & 8.0 & 14.0 \\
\textit{Ubc} & 7.0 & 7.0 \\
\textit{Eps8} & 5.0 & 5.0 \\
\textit{Rhoc} & 4.0 & 4.0 \\
\textit{Dynll1} & 4.0 & 6.0 \\
\textit{Zbtb32} & 4.0 & 4.0 \\
\textit{Foxm1} & 4.0 & 4.0 \\
\textit{Crip1} & 4.0 & --- \\
\textit{Mrpl52} & 4.0 & 5.0 \\
\textit{Sulf2} & --- & 4.0 \\
\end{tabular}
\end{center}
\end{table}

\begin{table}[H]
\caption{Top 20 Most Frequently Selected Genes (200 Initial Genes, 100 Repetitions)} \label{tab:top_genes_200}
\begin{center}
\begin{tabular}{lcc}
\textbf{Gene} & \textbf{Filter (\%)} & \textbf{Gradient (\%)} \\
\hline \\
\textit{Ccl3} & 60.0 & 49.0 \\
\textit{Sqstm1} & 60.0 & 46.0 \\
\textit{Ccl4} & 56.0 & 45.0 \\
\textit{Hsp90aa1} & 37.0 & 32.0 \\
\textit{Sdc4} & 30.0 & 22.0 \\
\textit{Tnfaip2} & 28.0 & 25.0 \\
\textit{Acod1} & 26.0 & 20.0 \\
\textit{Hmox1} & 22.0 & 16.0 \\
\textit{Esd} & 21.0 & 21.0 \\
\textit{Hspa8} & 15.0 & 11.0 \\
\textit{Sod2} & 13.0 & 8.0 \\
\textit{Gadd45b} & 11.0 & 11.0 \\
\textit{Rel} & 10.0 & --- \\
\textit{Odc1} & 9.0 & 14.0 \\
\textit{Crip1} & 8.0 & 9.0 \\
\textit{Tnip1} & 7.0 & --- \\
\textit{Ubc} & 6.0 & 8.0 \\
\textit{Txnrd1} & 6.0 & --- \\
\textit{2010111I01Rik} & 6.0 & --- \\
\textit{Chordc1} & 6.0 & 6.0 \\
\textit{Cyp51} & --- & 9.0 \\
\textit{Fosb} & --- & 8.0 \\
\textit{Btg2} & --- & 6.0 \\
\textit{Fcgr1} & --- & 6.0 \\
\end{tabular}
\end{center}
\end{table}

\begin{table}[H]
\caption{Top 20 Most Frequently Selected Genes (500 Initial Genes, 100 Repetitions)} \label{tab:top_genes_500}
\begin{center}
\begin{tabular}{lcc}
\textbf{Gene} & \textbf{Filter (\%)} & \textbf{Gradient (\%)} \\
\hline \\
\textit{Fas} & 35.0 & 2.0 \\
\textit{Ccl3} & 25.0 & 1.0 \\
\textit{Acod1} & 25.0 & 2.0 \\
\textit{Ccl4} & 24.0 & --- \\
\textit{Cdkn1a} & 23.0 & 2.0 \\
\textit{Abcg1} & 21.0 & 2.0 \\
\textit{Esd} & 17.0 & --- \\
\textit{Hsp90aa1} & 16.0 & --- \\
\textit{Sqstm1} & 15.0 & --- \\
\textit{Pim1} & 14.0 & --- \\
\textit{Traf1} & 12.0 & 3.0 \\
\textit{Tnfaip2} & 11.0 & --- \\
\textit{Cyb5a} & 10.0 & --- \\
\textit{Angpt2} & 10.0 & --- \\
\textit{Hmox1} & 10.0 & --- \\
\textit{Tnf} & 9.0 & --- \\
\textit{Dnaja1} & 9.0 & --- \\
\textit{Ehd1} & 9.0 & 2.0 \\
\textit{Sod2} & 8.0 & --- \\
\textit{Rel} & 8.0 & --- \\
\textit{Tnip1} & --- & 2.0 \\
\textit{Brd2} & --- & 2.0 \\
\textit{S100a6} & --- & 2.0 \\
\textit{Clec4e} & --- & 2.0 \\
\textit{Crybg1} & --- & 1.0 \\
\textit{Hpgds} & --- & 1.0 \\
\textit{Alas1} & --- & 1.0 \\
\textit{Susd3} & --- & 1.0 \\
\textit{Nfkbib} & --- & 1.0 \\
\textit{Lmna} & --- & 1.0 \\
\textit{Ppp3ca} & --- & 1.0 \\
\textit{Ptgs2} & --- & 1.0 \\
\textit{Acat2} & --- & 1.0 \\
\end{tabular}
\end{center}
\end{table}

\end{document}